\documentclass[sigconf]{acmart}
\usepackage{enumitem}
\usepackage{subcaption}

\AtBeginDocument{%
  }

\copyrightyear{2026}
\acmYear{2026}
\setcopyright{cc}
\setcctype{by-nc-nd}
\acmConference[SUI '26]{ACM Symposium on Spatial User Interaction}{October 10--11, 2026}{Bari, Italy}
\acmBooktitle{ACM Symposium on Spatial User Interaction (SUI '26), October 10--11, 2026, Bari, Italy}
\acmDOI{10.1145/3822518.3830043}
\acmISBN{979-8-4007-2812-9/2026/10}

\begin{document}

\title{``What Can I Do for You'': How Should AI Companions Provide Assistance to Players in Virtual Reality Games}

\author{Taiyu Zhang}
\affiliation{%
  \institution{KU Leuven}
  \city{Leuven}
  \country{Belgium}
}
\email{taiyu.zhang@kuleuven.be}

\author{Xinnian Zhao}
\affiliation{%
  \institution{KU Leuven}
  \city{Leuven}
  \country{Belgium}}
\email{xinnian.zhao@kuleuven.be}

\author{Adalberto L. Simeone}
\affiliation{%
  \institution{KU Leuven}
  \city{Leuven}
  \country{Belgium}}
\email{adalberto.simeone@kuleuven.be}


\begin{abstract}
  Recent advances in artificial intelligence (AI) have expanded the capabilities of non-player characters (NPCs), enabling them to perceive game states, perform in-game actions.
  {In immersive virtual reality (VR) games, such assistance is not limited to providing hints or interface-level support: an AI companion can appear as a co-present character, share the player’s spatial environment, and visibly act on game objects. This raises a design question for VR gameplay: how can AI companions best assist players while preserving their active participation in the virtual world?}
  To explore this question, we developed a VR puzzle game for Apple Vision Pro featuring an AI companion across four gameplay modes: no assistance, command-based assistance, discussion-based assistance, and autonomous agent play.
  A within-subjects study with 24 participants showed that AI assistance significantly reduced players’ workload. However, autonomous agent play, despite producing the lowest workload, substantially diminished player experience by reducing challenge, autonomy, immersion, and enjoyment. Qualitative analysis further showed that players evaluated the companion not only by its usefulness, but also by whether it felt like a tool, a teammate, or an integrated character in the game world. We categorised participants into four player types and summarised their expectations of AI companions. These findings provide design implications for AI companions as embodied participants in VR games.
\end{abstract}



\maketitle

\section{Introduction} 
Non-player characters (NPCs) are a central component of digital games, shaping how players interpret goals, interact with game worlds, and experience social presence. They can enhance immersion and player experience by providing realistic, responsive, and context-sensitive interactions~\cite{cai2024digital, yin2024press, kopel2018implementing, mooney2014rethinking, ochs2009simulation, saranya2023behaviors, zhu2019behavior}. With recent advances in AI technologies, NPCs have been equipped with increasingly sophisticated capabilities, enabling them to perform complex actions and execute strategies in game environments. In highly competitive settings, AI-controlled agents have matched or surpassed expert human performance in games like \emph{Dota 2}~\cite{dota2, berner2019dota}. 

\begin{figure}[t]
  \centering
  \includegraphics[width=\linewidth]{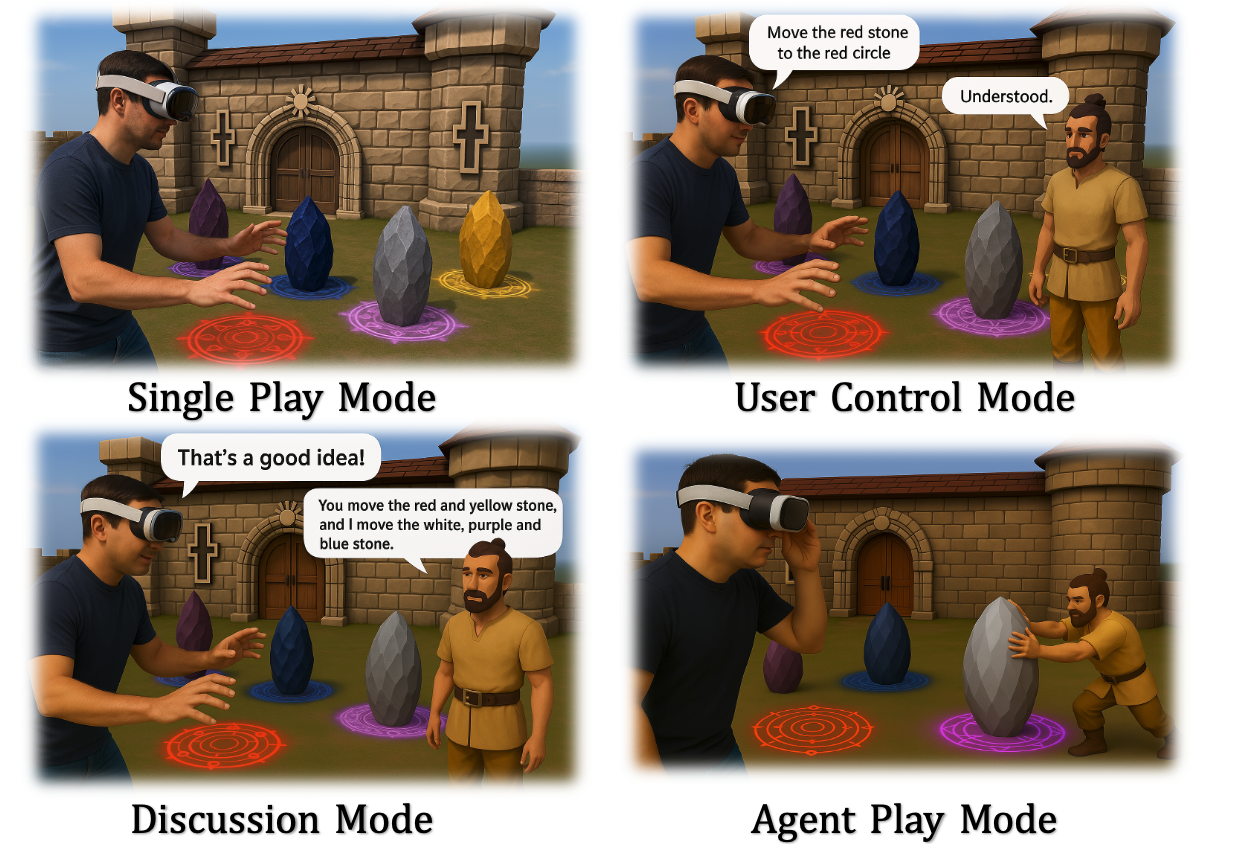}
  \Description[Four screenshots showing a player and an AI companion in a VR puzzle game]{A two-by-two grid of four game screenshots, each labelled with one gameplay mode. In all four, a player avatar wearing a virtual reality headset stands in a walled courtyard in front of a stone gate, surrounded by coloured magic stones and matching coloured circles marked on the ground. In Single Play Mode the player stands alone. In User Control Mode a bearded companion character has joined the player; a speech bubble from the player reads ``Move the red stone to the red circle'' and the companion replies ``Understood.'' In Discussion Mode the two exchange longer speech bubbles in which the companion proposes splitting the stones between them and the player answers ``That's a good idea!'' In Agent Play Mode the player stands back and watches while the companion moves the stones alone.}
  \caption{
      \textit{Break the Magic Circle} is a VR puzzle game featuring an interactive AI companion, 
      {designed to examine how different AI companion behaviours shape player workload and experience.}
      It includes four gameplay modes: (1) \textit{Single Play Mode}, where players solve puzzles independently; (2) \textit{User Control Mode}, where players issue direct commands to request assistance from the AI companion; (3) \textit{Discussion Mode}, where players and the AI companion discuss a strategy before executing actions together; and (4) \textit{Agent Play Mode}, where the AI companion autonomously completes the task without player interaction. This figure was produced using AI-assisted image generation and post-processing tools.
  }
  \label{fig:intro_fig}
\end{figure}

However, despite these advances, most NPCs in real-time games remain limited to scripted behaviours or rule-based systems, largely due to the practical challenges of training and deploying learning-based agents in diverse game contexts~\cite{choong2025support}.
More recently, large language models (LLMs) have opened new possibilities for player--NPC interaction by enabling natural language communication and context-aware responses~\cite{liu2023llm, shanahan2023role, akata2023playing, xu2023exploring, agashe2023evaluating}. Through conversational interaction, NPCs can interpret player intent, respond to player input, and provide assistance that adapts to ongoing gameplay. For example, NVIDIA has showcased AI-driven game characters in \emph{PUBG: BATTLEGROUNDS}~\cite{pubg}, supporting real-time interaction and action-level assistance~\cite{nvidia2024ace}. 

{Prior HCI games research identifies concrete player-facing value in such assistance. Studies of AI-supported onboarding and in-game LLM chatbots suggest that conversational and personalised guidance can help players access contextual information and understand unfamiliar game mechanics~\cite{choong2025support,lee2025development}. In cooperative play, an LLM assistant supported players' convergent and divergent thinking, illustrating how AI can contribute to collective problem solving~\cite{sidji2024human}. These benefits, however, are not devoid of issues. AI assistance can interfere with shared team mental models, while the allocation of initiative between the player and the AI can affect perceived control, warmth, and collaboration ~\cite{sidji2024human,lobo2024lead}. Prior work therefore suggests that the value of AI assistance depends not only on what the AI can do, but also on how its participation is integrated into play.}

{Accordingly, the central design question is not simply whether AI assistance can benefit players, but how assistance should be distributed between the player and the companion. It remains unclear how much assistance players want from NPC companions, how much control they wish to retain, and what forms of interaction they consider meaningful.}

{This question is important in VR games because assistance becomes spatially and bodily experienced. A companion in VR by virtue of being present in the environment together with the player, can provide actionable help beyond hints or information provided by a conventional game UI, it can discuss the current situation with the player, move around, and visibly perform actions that the player might otherwise execute themselves. As a result, AI assistance may reduce cognitive and physical workload while also changing the player’s sense of agency, spatial involvement, and social presence. Understanding this balance is therefore important for designing AI companions that support players without displacing them from active participation in immersive gameplay.}

To explore this question, we implemented an AI NPC companion in a VR puzzle game, integrating game-state awareness with real-time speech input, as shown in Fig.~\ref{fig:intro_fig}. 
We designed four gameplay modes that vary in the 
{forms of AI companion participation} and degree of player interaction:

\begin{itemize}
\item \textbf{Single Play Mode:} Players solve the puzzle independently without assistance from the AI companion.
\item \textbf{User Control Mode:} Players retain control over the task and can issue direct commands for the AI companion to perform specific in-game actions.
\item \textbf{Discussion Mode:} Players and the AI companion first discuss a strategy, after which they execute the plan together through coordinated actions.
\item \textbf{Agent Play Mode:} The AI companion autonomously completes the puzzle, providing full assistance without player interaction.
\end{itemize}

We first conducted a pilot study to evaluate the AI NPC framework’s understanding and action capabilities during player communication and in-game interaction. Successively, we conducted a within-subjects user study with 24 participants who experienced the game in four different modes to investigate the impact of the extent of assistance and interaction provided by an AI companion. After each mode, participants completed the Mini-PXI~\cite{haider2022minipxi} to assess player experience and the NASA-TLX~\cite{hart1988development} to assess workload. After the study, we conducted semi-structured interviews to understand participants' gaming habits, their experiences with the four modes, and their feedback on the intelligent NPC system.

The quantitative analysis showed that the AI companion effectively reduced players’ workload during gameplay. However, fully automated assistance without player interaction reduced overall game experience, despite producing the lowest workload. This finding suggests that players do not simply prefer more assistance; rather, they value maintaining agency and engaging in meaningful interaction with NPC companions. To further examine these expectations, we categorised participants into four groups according to their everyday gaming habits from interview responses. The qualitative findings showed that outcome-oriented players preferred efficient, task-focused exchanges, whereas narrative-oriented players favoured richer and more expressive interactions. Competitive players' preferences depended on whether they prioritised efficiency, control, or cooperation, while socially oriented players emphasised a sense of connection with the companion.


{Overall, this work makes two contributions. First, we present and evaluate an embodied AI NPC companion for VR gameplay that combines speech interaction, game-state awareness, and visible in-world action execution. Rather than functioning as an external hint system, the companion participates in the same virtual space as the player and can directly affect shared game objects. Second, we empirically compare four practical interaction designs that redistribute planning, execution, initiative, and player activity between the player and the AI companion. The findings show that AI assistance can reduce workload, but that autonomous assistance without player interaction can weaken players’ sense of active participation and game experience. These results provide design implications for future LLM-supported NPCs and AI companions in immersive VR games.}

\section{Related Work}

\begin{figure*}[!tbp]
    \centering
    \includegraphics[width=0.9\textwidth]{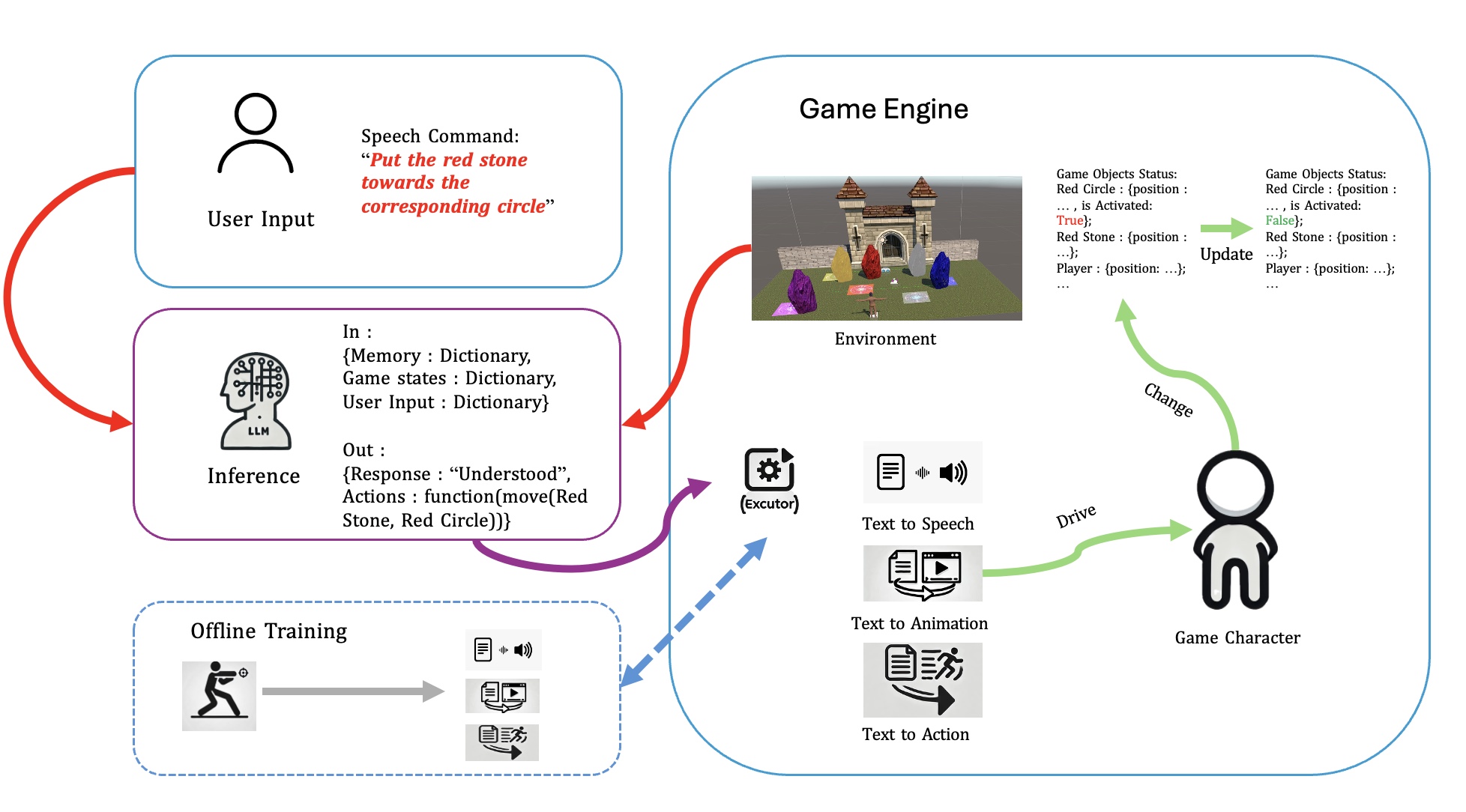} 
    \Description[Block diagram linking user speech, a language model, an executor, and the game engine]{A block diagram with four labelled panels connected by coloured arrows. At the upper left, a ``User Input'' panel shows the speech command ``Put the red stone towards the corresponding circle.'' A red arrow carries it into an ``Inference'' panel, which lists a dictionary-shaped input of memory, game states and user input, and a matching output containing a spoken response and a function call such as move(Red Stone, Red Circle). A purple arrow passes that output to an ``Executor'' block feeding three modules labelled Text to Speech, Text to Animation and Text to Action. These drive a game character inside a large ``Game Engine'' panel on the right, which also contains a rendered view of the puzzle environment and a listing of game-object states; a green arrow shows the character's action changing an object's activation flag from True to False and updating the listing. A dashed blue arrow links a separate ``Offline Training'' panel at the lower left to the three conversion modules.}
    \caption{Overview of the AI NPC framework. The system converts player speech and game-state data into textual context for language-model reasoning, then parses structured outputs into spoken responses and in-game actions.}
    \label{fig:framework}
\end{figure*}

\subsection{Technical Evolution of Intelligent NPCs}
The concept of non-player characters (NPCs) can be traced back to the earliest digital games. Pioneering titles such as \emph{OXO} (1952), \emph{Tennis for Two} (1958), and \emph{Spacewar!} (1962) featured simple rule-based opponents that responded to player inputs in highly constrained ways~\cite{oxo,tennis_for_two,spacewar}. Later, role-playing games such as \emph{Ultima V–VII} introduced scheduled routines and scripted dialogues, providing an early glimpse of life-simulation style agency~\cite{ultima_v,ultima_vii}. 

Early NPCs were typically controlled by hand-crafted scripts or finite-state machines, resulting in limited, repetitive, and predictable behaviours~\cite{yannakakis2014panorama}. Such approaches offered only shallow interactivity and could break immersion once players recognised their patterns. With the rise of machine learning, reinforcement learning enabled NPCs to adapt dynamically to changing contexts and optimise strategies through trial and error. A milestone example is OpenAI Five, whose agents competed with professional players in \emph{Dota 2}~\cite{berner2019dota}.  Advances in deep learning and multimodal modelling have also been explored in game contexts, for example by using vision-language models to guide agents through partial gameplay sequences and generate context-sensitive behaviours in complex environments such as \emph{Black Myth: Wukong}~\cite{black_myth_wukong, chen2024can, tan2024towards}. Together, these developments mark a transition from static, rule-based agents toward adaptive NPCs capable of dialogue, strategic reasoning, and collaboration.

More recently, LLMs have further expanded NPC capabilities by supporting context-aware dialogue, role-consistent responses, and memory across interactions~\cite{guo2025deepseek, achiam2023gpt, park2023generative, dubey2024llama}. Practical toolkits such as \emph{CUIfy the XR} provide open-source infrastructure for embedding LLM-powered conversational agents in XR environments, supporting multi-agent voice interaction with reduced latency~\cite{buldu2025cuify}. Although recent efforts such as NVIDIA’s ACE\footnote{ACE: Avatar Cloud Engine, a framework for building real-time, generative AI-powered NPCs by NVIDIA.} demonstrate feasibility of real-time conversational NPCs, they often remain platform-level demonstrations rather than studies of how players experience different forms of NPC assistance during gameplay. Our work addresses this gap by embedding an AI NPC companion within a VR puzzle game and empirically examining how different assistance and interaction modes affect player workload and experience.

\subsection{Player Expectations of AI NPCs}

While prior work has emphasised technical advances in AI-driven NPCs, less is known about how players perceive, use, and evaluate such NPCs during gameplay. Much of the existing literature focuses on improving NPC intelligence, autonomy, or task performance, often treating increased capability as a proxy for improved player experience. However, this assumption overlooks the fact that players’ expectations of NPCs are shaped not only by functional competence, but also by how NPCs communicate, behave, and situate themselves within the gameplay context.

Recent studies highlight the importance of multimodal cues, including voice, gaze, gestures, and body movement, in shaping players' perceptions of NPCs, particularly in immersive VR environments~\cite{simeone2015substitutional, yin2024press, cheng2017teaching, elliman2016virtual}. These cues are important for establishing social presence, intelligibility of intent, and perceived responsiveness, which in turn affect whether interactions with NPCs feel natural or artificial. Consequently, intelligence alone is insufficient to ensure engaging or satisfying NPC interactions.

Prior work on player experience has shown that enjoyment depends on factors such as challenge, control, feedback, immersion, and social interaction~\cite{sweetser2005gameflow}. {Self-determination theory similarly identifies autonomy and competence as core drivers of game enjoyment~\cite{ryan2006motivational}, and player typology research suggests that these needs are weighted differently across player groups~\cite{bartle1996hearts, yee2006motivations}. Beyond games, research on human--automation interaction has long shown that increasing automation lowers workload but can reduce engagement and involvement~\cite{parasuraman2000model, bainbridge1983ironies}.} However, AI NPC assistance raises a distinct question: assistance can reduce task demands while simultaneously changing player agency, control, and the social form of interaction with the companion. This motivates our focus on how different AI NPC assistance modes shape workload and player experience in VR gameplay.

Empirical evaluations further suggest a mismatch between technical capability and experiential quality. Although LLM- or AI-driven NPCs are often rated highly in behavioural diversity or perceived intelligence, they tend to fall short in emotional depth, personality coherence, and long-term consistency~\cite{korkiakoski2025empirical}. This gap suggests that players may evaluate NPCs along dimensions that are not captured by traditional AI benchmarks, including appropriateness of assistance, social sensitivity, and alignment with player intent.


{Building on this body of work, our study shifts the focus from how intelligent NPCs can become to how AI companions should participate in immersive gameplay. In VR, assistance is not only a matter of providing better information or more capable automation. Because the companion can be embodied as a co-present character and can visibly act on shared game objects, assistance also shapes the player’s spatial agency, physical involvement, and sense of social presence. We therefore investigate how different AI companion interaction designs affect workload, player experience, and engagement in a VR game, aiming to better align AI-driven NPC behaviours with players’ expectations in immersive gameplay.}

\section{AI NPC Framework}


As shown in Fig.~\ref{fig:framework}, we implemented an AI NPC framework that connects player input, game-state data, language-model reasoning, and in-game action execution. 
Rather than relying on a visual model to infer the game world only from rendered frames, the framework communicates directly with the game engine to access task-relevant state variables. 
This allows the NPC to reason over the current puzzle state and execute supported actions during gameplay. The framework comprises three layers: Perception, Cognitive, and Executor.

\subsection{Perception Layer} 
The Perception Layer converts two input streams into text descriptions for language-model inference: player speech and game-state data. For player speech, we apply Whisper to transcribe voice commands~\cite{radford2023robust}. Game-state data are obtained through direct queries to the Unity game engine, including task-relevant information such as object positions and puzzle states. Existing approaches often rely on vision-language models to process gameplay frames~\cite{chen2024can,tan2024towards,nvidia2024ace}, but frame-based perception can be unreliable in complex scenes~\cite{bordes2024introduction} and limits the NPC to the player's visual perspective. Direct engine communication avoids this viewpoint constraint and provides the Cognitive Layer with structured game-state information from game objects in the engine. The shared data include symbolic state variables such as object positions, object types, and interaction states. For example, in the puzzle task, the system retrieves the positions of stones, target locations, doors, and the player character from Unity, allowing the LLM to reason about actions such as moving a stone onto a pressure plate to unlock a door. Since the Cognitive Layer operates on structured object states rather than game-specific visual input, this framework can potentially generalise to other games.

\subsection{Cognitive Layer} 
The Cognitive Layer passes these text descriptions to GPT-4~\cite{achiam2023gpt}\footnote{OpenAI API model: \texttt{gpt-4-turbo-2024-04-09}.}, which is prompted to act as the AI NPC. The prompt contains the player's latest utterance, prior interaction context, and the current game-state description. The model generates structured outputs specifying a spoken response and, when appropriate, one or more supported game actions. Using structured outputs allows the Executor Layer to map model decisions to validated action commands rather than interpreting free-form text.

\subsection{Executor Layer} 
The Executor Layer parses and validates the structured outputs, then maps them to game responses. In our prototype, this includes text-to-speech synthesis for spoken feedback and two game-engine actions: moving to a target stone and pushing a stone. These actions are intentionally constrained to the mechanics required by the puzzle, reducing execution ambiguity while supporting the assistance behaviours needed for the study.

We integrated the framework into a first-person VR puzzle game, \emph{Break the Magic Circle}, built with Unity 6 for Apple Vision Pro. Players must push “magic stones” onto colour-matched circles to open a door. The same AI NPC framework is used across all study conditions, while the NPC's initiative and interaction flow vary by mode: no assistance in Single Play, command-triggered action assistance in User Control, strategy discussion and joint execution in Discussion, and autonomous task completion in Agent Play. {To support reproducibility, the prompt template and the structured output schema used by the Cognitive Layer are provided in Appendix~\ref{app:prompt}.}
\section{Study Design and Pilot Study}


{To investigate players’ expectations regarding AI companions in gameplay, we designed four interaction modes that represent practical configurations of AI companion participation. These modes differed in how planning, execution, initiative, interaction, and player activity were distributed between the player and the AI companion:}

\begin{enumerate}
    {
    \item \textbf{Single Play Mode}: The player completes all tasks independently without AI assistance.
    \item \textbf{User Control Mode}: The AI companion assists only when explicitly commanded. The player remains responsible for game analysis and strategy, while the AI companion executes delegated actions.
    \item \textbf{Discussion Mode}: The AI companion collaborates with the player to jointly formulate and execute strategies, providing support for both decision-making and task execution.
    \item \textbf{Agent Play Mode}: The AI companion autonomously manages all tasks, requiring no intervention from the player.
    }
\end{enumerate}

Based on these conditions, we formulated the following hypotheses:

\begin{enumerate}[label=\textit{H\arabic*}, leftmargin=*]
    \item AI companions reduce players' workload during gameplay.
    \item {Autonomous assistance without meaningful interaction diminishes players' gaming experience.}
    \item Moderate assistance improves gaming experience, but too much help degrades it.
\end{enumerate}

\begin{figure}[!t]
    \centering
    \includegraphics[width=0.95\linewidth]{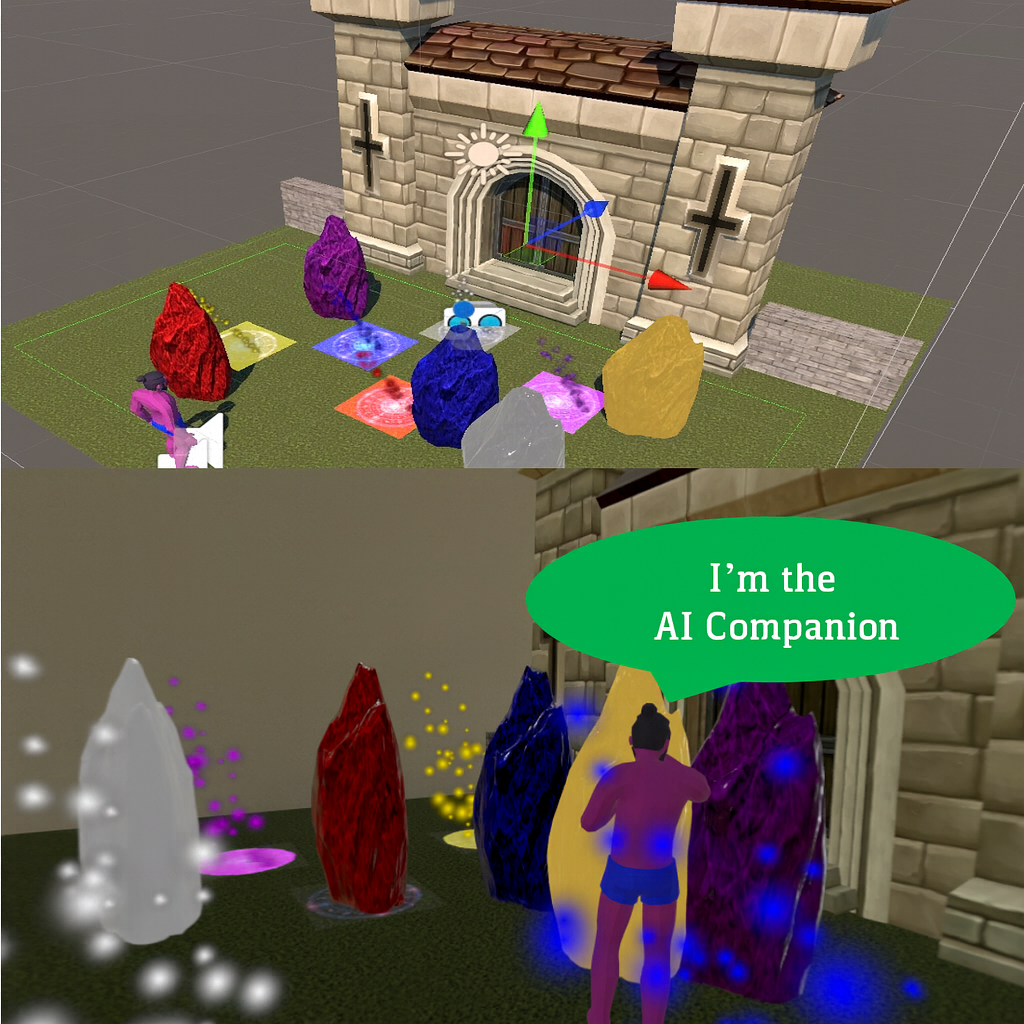}
    \Description[Top: the puzzle level seen from above in the editor. Bottom: the same scene at player eye level in the headset.]{Two stacked images of the same environment. The upper image shows the level from an elevated angle: a stone gateway set into a wall with a glowing portal at its centre, standing on a grass platform. Nine faceted crystal stones in red, purple, blue, yellow and white are scattered across the platform, and flat circles in matching colours are marked on the ground in front of the gate. The lower image shows the environment from the player's eye level, lit far more darkly so that the stones glow against the background. The companion character stands beside the player next to a large blue stone, with a green speech bubble reading ``I'm the AI Companion.''}
    \caption{Game scene and player perspective in \emph{Break the Magic Circle}. Top: puzzle layout in the Unity game engine. Bottom: first-person view on Apple Vision Pro.}
    \label{fig:gamescene}
\end{figure}

We used the puzzle game described in the previous section to test these hypotheses. Five coloured ``magic circles'' on the ground controlled the opening and closing of a door in the virtual environment. Players had to destroy the circles by moving colour-matched ``magic stones'' onto them; once all circles were eliminated, the door opened and the game was completed. Figure~\ref{fig:gamescene} shows the puzzle layout and the first-person view experienced by participants. To mitigate motion sickness, the game was deployed on Apple Vision Pro with a gesture-based interface that allowed players to push stones without physical controllers.

We conducted a pilot study with 10 volunteers to assess the stability and usability of the AI companion system. Each volunteer experienced all four modes during the pilot study. We recorded interactions between volunteers and the AI companion, and conducted a short interview after each session to collect system feedback. Across all volunteers, 48 commands were issued in User Control Mode. Of these, 47 commands were successfully recognised, and 46 were executed correctly on the first attempt, yielding an execution accuracy of $96\%$ and an $F_1$-score of $0.98$\footnote{The $F_1$-score is defined as $F_1 = \tfrac{2 \cdot \text{Precision} \cdot \text{Recall}}{\text{Precision} + \text{Recall}}$, 
where Precision = $\tfrac{TP}{TP+FP}$ and Recall = $\tfrac{TP}{TP+FN}$. Here, $TP$ denotes correctly executed commands, $FP$ mis-executed commands, and $FN$ unretrieved commands.}.

Interaction latency was approximately 1--3 seconds in User Control Mode and 3--5 seconds in Discussion Mode, reflecting the longer conversational exchanges in Discussion Mode. To reduce the potential impact of waiting delays on user experience, we introduced a status icon in the upper-right corner of the field of view to indicate the current state of the AI companion.
In the pilot study, participants did not report notable frustration related to waiting times. This feedback suggested that the visual status cue helped make the system state more transparent and predictable during delayed responses.

\section{User Study}

To explore players' expectations of collaborative AI NPCs, we adopted a within-subjects design with mode order counterbalanced using a Latin square. Table~\ref{tab:latin} presents the sequence of game modes assigned to each group. All participants completed the study in person in our laboratory, where we provided the Apple Vision Pro and assisted them with eye- and hand-tracking calibration before the study. Before beginning, each participant was assigned to a counterbalancing group and signed an informed consent form detailing the experimental procedures and data-use policy.  

At the start of the session, participants reported their gender and age in the application. They then experienced the four game modes in the assigned order. {In Single Play Mode, the player manually pushed each stone to its target circle to unlock the door. In User Control Mode, the player activated the AI companion using the wake phrase ``Hey agent'' and then issued verbal commands, which the AI companion executed. In Discussion Mode, the AI companion first proposed a strategy; after the player confirmed the plan, the player and AI companion cooperatively completed the task. In Agent Play Mode, the AI companion pushed all required stones to their corresponding circles.} All interactions with the AI companion were conducted in English. After completing each mode, participants completed the MiniPXI~\cite{haider2022minipxi} and NASA-TLX~\cite{hart1988development} questionnaires in the application.

Following gameplay, participants took part in a semi-structured interview, which was audio-recorded with their consent. The interview focused on three topics: (1) general gaming habits, introduced with the prompt \textit{``What kind of player do you consider yourself?''}. If participants could not answer directly, we followed up with questions about their daily gaming habits and prior gaming experiences, such as \textit{``Which games do you typically play in your daily life?''} and \textit{``What aspects attract you most when playing games?''}; (2) experiences during study, with questions such as \textit{``How did you feel in each mode, and what were your impressions of the AI companion?''}; and (3) feedback on the system.

Participants were recruited through social media advertisements and physical posters distributed around the university. We recruited 24 participants, including 11 males, 12 females, and one participant who identified as other (mean age = 28.71 years, SD = 6.97, range = 19--45), all of whom reported prior video game experience. The participants were divided into four counterbalancing groups of six, corresponding to the Latin-square design. Each session lasted approximately 30 minutes. 
The study protocol, covering both the pilot study and the main user study, was approved by the Social and Societal Ethics Committee (SMEC) of KU Leuven through KU Leuven's Privacy and Ethics platform (PRET, approval no.~G-2025-8976), in compliance with the GDPR. All participants provided written informed consent.

\begin{table}[h]
\centering
\caption{Latin-square ordering of game modes.}
\label{tab:latin}
\begin{tabular}{lllll}\toprule
Group & 1st & 2nd & 3rd & 4th\\\midrule
A & Single & Agent & User & Discussion\\
B & Agent & Discussion & Single & User\\
C & User & Single & Discussion & Agent\\
D & Discussion & User & Agent & Single\\
\bottomrule
\end{tabular}
\end{table}

\section{Results}

We collected both quantitative and qualitative data during the user study. Quantitative data were analysed following the original questionnaire guidelines, and interview recordings were transcribed and reviewed by two researchers fluent in English.

\subsection{Quantitative Results}

\begin{figure}[htbp]
\Description[Box plot and six line charts showing workload falling from single play to agent play, except for reverse-coded performance]{The figure has two panels. The upper panel is a box plot with four boxes labelled single, user control, discussion and agent play on the horizontal axis and NASA-TLX score from 0 to 20 on the vertical axis, with individual participant scores overlaid as translucent dots. The single condition has the highest and widest distribution, centred near 7 and spanning roughly 3 to 12; user control and discussion are lower and similar to each other, both centred near 5; agent play is lowest and narrowest, centred near 3.7. All four distributions sit in the lower half of the available scale. The lower panel is a grid of six small line charts, one per subscale, each plotting the mean with a shaded 95\% confidence band across the same four modes. Mental demand rises slightly from about 9 to 12 between single and user control, then falls to about 1 at agent play. Physical demand falls steeply from about 15 to about 1, and effort follows a similar path from about 14 to about 1. Reverse-coded performance is the only chart that rises, staying near 1.5 for the first three modes before climbing sharply to about 14 at agent play. Temporal demand stays close to 0 throughout and frustration stays flat between 3.5 and 5, both with confidence bands that overlap across all four modes.}
    \centering
    \begin{subfigure}{\linewidth}
        \centering
        \includegraphics[width=\linewidth]{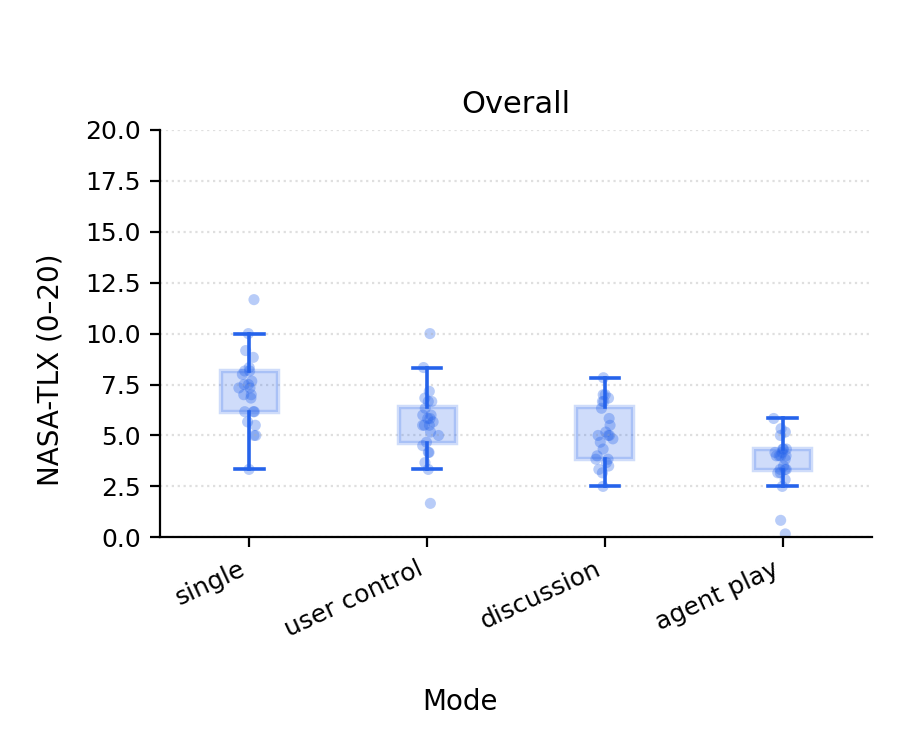}
        \caption{Overall workload}
        \label{fig:workload_box}
    \end{subfigure}

    \begin{subfigure}{\linewidth}
        \centering
        \includegraphics[width=\linewidth]{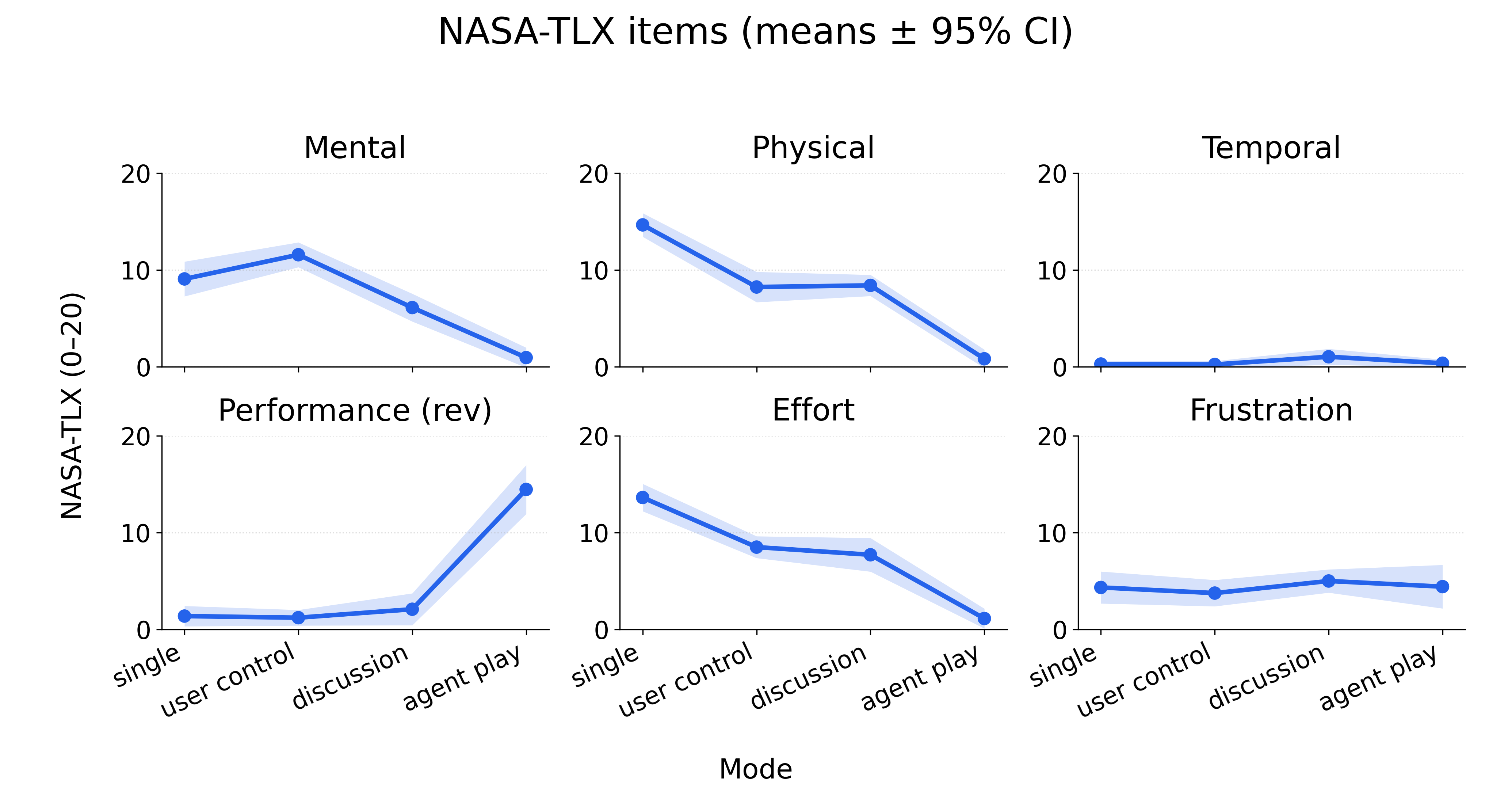}
        \caption{NASA--TLX subscales}
        \label{fig:workload_line}
    \end{subfigure}

    \caption{NASA--TLX workload across gameplay modes}
    \label{fig:workload}
\end{figure}

A total of 24 participants completed the quantitative questionnaires.
{We analysed workload using Raw NASA–TLX scores. In our in-game questionnaire, each NASA–TLX subscale was rated on a 21-point scale from 0 (very low) to 20 (very high), and the overall workload score was calculated as the unweighted mean of the six subscales, with Performance reverse-coded.}
Game experience was analysed using MiniPXI construct scores. Following the MiniPXI validation study, we analysed MiniPXI scores using repeated-measures ANOVAs, with Greenhouse--Geisser corrections applied when sphericity was violated. Holm-corrected paired comparisons were used for post-hoc tests.

Figure~\ref{fig:workload} summarises the NASA--TLX workload results across gameplay modes. A repeated-measures ANOVA revealed a significant main effect of gameplay mode on overall workload, $F(3,69)=20.81$, $p<.001$, $\eta^2_p=.475$. Single Play resulted in the highest workload ($M=7.23$, $SD=1.77$), followed by User Control ($M=5.59$, $SD=1.68$) and Discussion ($M=5.06$, $SD=1.45$), while Agent Play yielded the lowest workload ($M=3.69$, $SD=1.27$). Holm-corrected post-hoc tests showed that Agent Play produced significantly lower workload than all other modes, and that Single Play produced significantly higher workload than both User Control and Discussion. No significant difference was found between User Control and Discussion. {Exact statistics for all pairwise comparisons, including non-significant ones, are reported in Appendix~\ref{app:posthoc}.}

Table~\ref{tab:workloadsubscale} reports the repeated-measures ANOVA results for each NASA--TLX subscale. Subscale analyses showed significant effects of gameplay mode on Mental Demand, Physical Demand, reverse-coded Performance, and Effort, but not on Temporal Demand or Frustration. Agent Play consistently reduced mental, physical, and effort-related workload, but was associated with lower perceived performance after reverse coding. Single Play produced the highest physical demand and effort. The lack of a significant difference between User Control and Discussion suggests that reduced manual effort may have been offset by the additional interaction required for discussion.

\begin{table}[t]
\centering
\caption{Repeated-measures ANOVAs on NASA--TLX subscales.}
\label{tab:workloadsubscale}
\begin{tabular}{lccc}
\toprule
Subscale & $F$ & $p$ & $\eta^2_p$ \\
\midrule
Mental Demand      & 44.69 & $<.001$ & .660 \\
Physical Demand    & 90.67 & $<.001$ & .798 \\
Performance (rev.) & 65.82 & $<.001$ & .741 \\
Effort             & 61.44 & $<.001$ & .728 \\
Temporal Demand    & 2.28  & .136    & .090 \\
Frustration        & 0.40  & .682    & .017 \\
\bottomrule
\end{tabular}
\end{table}

\begin{figure}[htbp]
\Description[Eleven line charts of player-experience constructs, nearly all dropping sharply at agent play]{A grid of eleven small line charts, one per MiniPXI construct, each plotting the mean with a shaded 95\% confidence band across the four modes in the order single, user control, discussion, agent play, on a vertical axis from 1 to 7. Most charts share the same shape: values stay between roughly 5 and 6.5 for the first three modes, then fall steeply at agent play, reaching about 2 for audiovisual appeal, challenge, curiosity, meaning and enjoyment, and about 1.7 for autonomy, the lowest point anywhere in the grid. Immersion and mastery differ in starting lower at single play, around 4 and 4.7, peaking at user control, and falling to about 3 and 2.5. Clarity of goals is the flattest chart, declining only from about 6 to about 4.8, and is the only construct whose confidence bands overlap substantially across all four modes.}
    \centering
    \includegraphics[width=\linewidth]{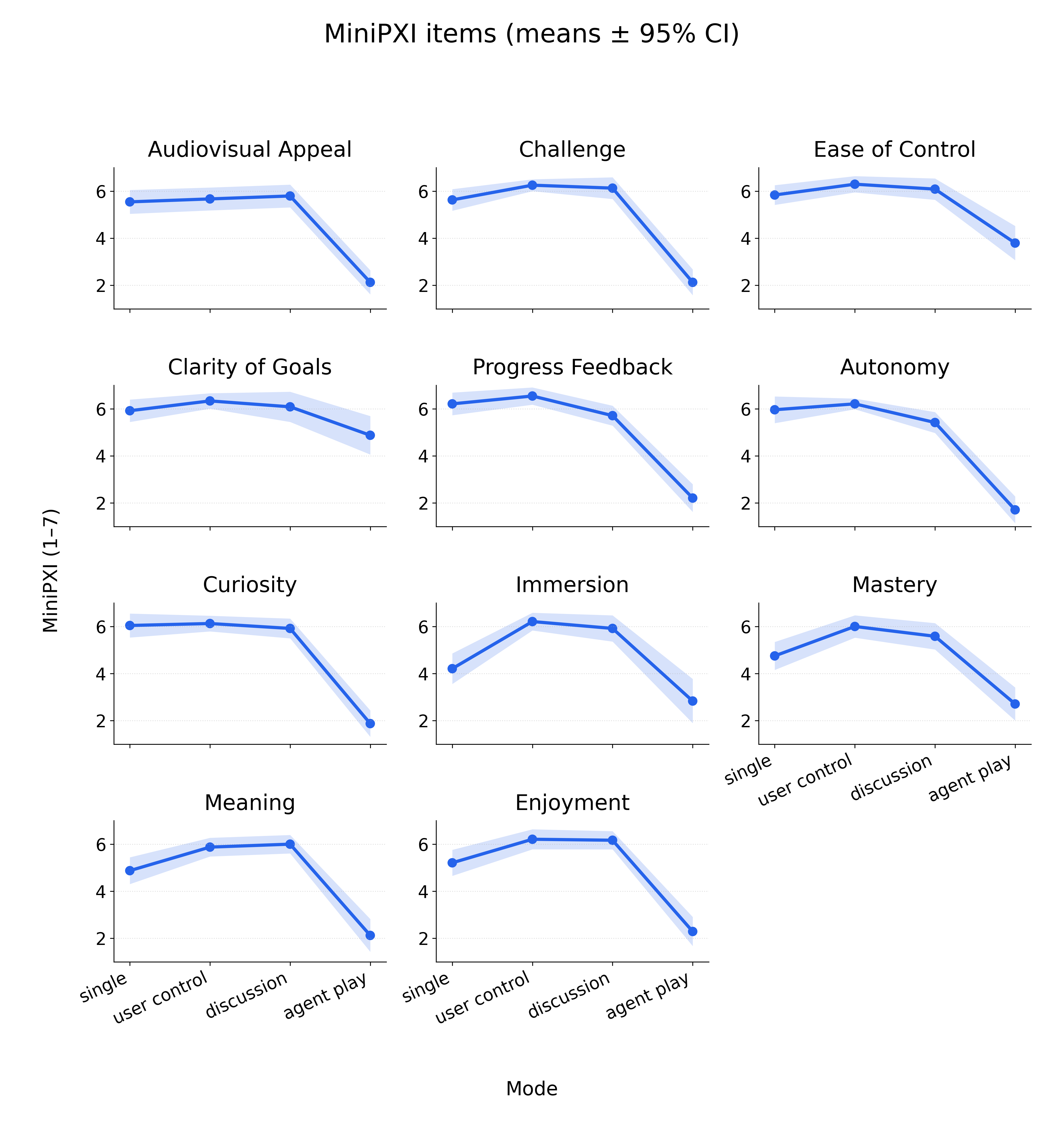}
    \caption{MiniPXI player experience scores across gameplay modes.}
    \label{fig:gameexperience}
\end{figure}

MiniPXI results also showed strong effects of gameplay mode on player experience, as shown in Figure~\ref{fig:gameexperience}. Composite MiniPXI scores differed significantly across modes, with User Control receiving the highest overall score ($M=6.16$, $SD=0.46$), followed by Discussion ($M=5.89$, $SD=0.64$), Single Play ($M=5.47$, $SD=0.65$), and Agent Play ($M=2.61$, $SD=1.18$). A repeated-measures ANOVA on the MiniPXI total score confirmed a significant main effect of mode, $F(2.12,48.82)=103.70$, $p_{GG}<.001$, $\eta^2_p=.818$. {Holm-corrected pairwise comparisons for the composite score are reported in Appendix~\ref{app:posthoc}.}

Table~\ref{miniPXI} reports the repeated-measures ANOVA results for MiniPXI constructs. At the construct level, repeated-measures ANOVAs showed significant effects of gameplay mode across all MiniPXI constructs. Agent Play was rated significantly lower than the other modes on most constructs, including audiovisual appeal, challenge, ease of control, progress feedback, autonomy, curiosity, mastery, meaning, and enjoyment. User Control and Discussion generally received the highest ratings, particularly for autonomy, immersion, meaning, and enjoyment. The effect on Clarity of Goals was weaker than the other constructs and should be interpreted cautiously, as it was significant in the parametric analysis but not in the non-parametric sensitivity check. 
{Overall, these results indicate that the Agent Play condition, which combined autonomous task execution with the absence of player interaction, produced the lowest workload but substantially diminished players’ sense of agency and game experience.}

\begin{table}[t]
\centering
\caption{Repeated-measures ANOVAs on MiniPXI constructs.}
\label{miniPXI}
\begin{tabular}{lccc}
\toprule
Construct & $F$ & $p$ & $\eta^2_p$ \\
\midrule
AV Appeal & 54.43 & $<.001$ & .703 \\
Challenge & 83.15 & $<.001$ & .783 \\
Ease Ctrl & 22.15 & $<.001$ & .491 \\
Clarity   & 4.99  & .014    & .178 \\
Progress  & 77.16 & $<.001$ & .770 \\
Autonomy  & 84.66 & $<.001$ & .786 \\
Curiosity & 85.85 & $<.001$ & .789 \\
Immersion & 24.22 & $<.001$ & .513 \\
Mastery   & 26.46 & $<.001$ & .535 \\
Meaning   & 49.63 & $<.001$ & .683 \\
Enjoyment & 57.11 & $<.001$ & .713 \\
\bottomrule
\end{tabular}
\end{table}

Overall, the quantitative results indicate that AI assistance can effectively reduce workload, but fully automated assistance without interaction can negatively affect players' game experience.

\subsection{Qualitative Results}
We first conducted a thematic analysis of player profiles, categorising participants into four groups according to their engagement with games, following the approach of Gerling et al.~\cite{gerling2020virtual}. A total of 51 unique codes were identified from the interview transcriptions and organised into three themes aligned with the interview structure: player profiles, preferences for AI companions, and system feedback. 
We then summarised players' game experiences and system feedback through content analysis, following the method of Zhang et al.~\cite{focusAgent}.

\subsubsection{Player Profiles}
Thematic analysis of the semi-structured interviews highlighted participants' motivations and habits related to gameplay. When asked the guiding question \textit{``What kind of player do you consider yourself?''}, most participants initially struggled to self-describe and required further prompts regarding their gaming habits and prior experiences. Through this process, each participant articulated a dominant aspect of their play style, which we used to classify participants into four categories.

\begin{itemize}[leftmargin=*]
    \item \textbf{Task-Oriented Players (TP; 9; P3, P4, P9, P16, P17, P19, P20, P21, P23)} focused primarily on task completion and outcomes. Several described limited interest in games, often playing only for practical goals or specific tasks. For instance: \textit{``I enjoyed games when I was a kid. But I don't spend time on games now.'' (P4)}. One participant emphasised their lack of gaming proficiency: \textit{``I rarely play games because I'm not good at it.'' (P16)}.
    \item \textbf{Competitive Players (CP; 5; P1, P2, P5, P15, P22)} emphasised competition and cooperation in team-based contexts. As one participant put it: \textit{``I always play competitive games.'' (P1)}.
    \item \textbf{Narrative Players (NP; 8; P6, P7, P8, P11, P12, P13, P18, P24)} were primarily motivated by story and in-game interaction. They highlighted narrative elements as the most engaging aspect of gameplay: \textit{``The story of the game is the most attractive part for me. I enjoy the details and interactions during the game.'' (P12)}.
    \item \textbf{Social Players (SP; 2; P10, P14)} valued social interaction over core gameplay. Both players reported that they typically engage in gaming only when accompanied by friends. \textit{``I often played games with my friends and enjoyed the time we spent together more than the game itself.'' (P10)}. 
\end{itemize}

{We note that this classification was conducted post hoc from self-reported habits and that the resulting groups were unbalanced in size, with only five competitive and two social players. We therefore use the player types as an analytic lens for organising the qualitative findings, and treat type-specific observations in the following sections as exploratory rather than as generalisable claims.}

\subsubsection{Preferences for AI Companion}
When discussing the four game modes, participants expressed distinct preferences regarding the AI companion. Among the 24 participants, 10 preferred User Control Mode (5 TP, 3 CP, 1 NP, 1 SP), 11 favoured Discussion Mode (3 TP, 2 CP, 5 NP, 1 SP), and 3 selected Single Play Mode (1 TP, 2 NP); no participant selected Agent Play Mode as their preferred mode. We then analysed the reasoning behind these choices. 

In Agent Play Mode, participants appreciated that the companion was helpful but criticised the lack of interaction. As one noted: \textit{``I don't know what I should do in this mode. It's more like an experience or guidance than a game.'' (P8)}. 

In Discussion Mode, participants were impressed by the agent's reasoning, reliable actions, and understanding of input. One participant described it as: \textit{``This process feels a bit like a negotiation, and I enjoy it.'' (P1)}. Some TPs valued the step-by-step reasoning but found the extended conversations time-consuming: \textit{``I could have finished the game if there was no discussion.'' (P9)}. Most CPs enjoyed the interactivity, although some felt that the game was too easy to justify such long conversations. Most NPs praised the conversations and felt that longer dialogue enhanced their immersion during gameplay. However, some criticised the system's conversational limitations: \textit{``When I tried to talk about something else, the companion did not understand.'' (P4)}. SPs valued the sense of connection but were frustrated by long response times: \textit{``The discussion process made me feel more connected, but the long waiting time weakened this connection.'' (P10)}.

In User Control Mode, participants generally found the companion accurate and responsive: \textit{``The AI companion in this mode is really smart. I like working together with that guy.'' (P2)}. However, the primary complaint in this mode was that the companion felt more like a tool than an integral part of the game, which reduced enjoyment for some participants. \textit{``It's more like a tool than a part of the game.'' (P13)}. 5 TPs and 3 CPs preferred this mode because it offered an efficient way to complete the task. \textit{``It's good that I don't need to do a lot of boring stuff, and it's interesting to see an AI helping me.'' (P3)}  \textit{``He is a good soldier to me.'' (P2)}.  Most NPs appreciated the responsiveness but wanted deeper integration with the game world: \textit{``I still hope this character could appear more authentic--for example, by adding backstory or more facial expressions.'' (P6)}. SPs highlighted real-time responsiveness as a source of engagement: \textit{``The companion felt responsive and reacted to my actions dynamically, like a real friend.'' (P14)}.

In Single Play Mode, a few participants preferred playing alone, focusing solely on VR gameplay and sometimes perceiving the companion as intrusive: \textit{``I enjoy the actions in the game and don't really need help from the agent.'' (P9)}. However, others felt that the AI companion added enjoyment to the game: \textit{``This AI companion brings additional fun to this game.'' (P3)}.

\subsubsection{System Feedback}
Overall, participants reported that the AI companion functioned effectively. The main criticism concerned latency, particularly in Discussion Mode. Player types also shaped participants' expectations of the companion. Most TPs expressed interest in more challenging tasks involving the intelligent companion: \textit{``I can't wait to play with this companion on more challenging tasks.'' (P9)}. 3 CPs desired more dynamic companion behaviours, with one noting interference: \textit{``When I tried to move a stone, the AI moved another stone that blocked me'' (P2)}. Most NPs and SPs expected the AI companion to feel more embedded within the game world: \textit{``I hope this NPC could be part of the game world rather than a tool.'' (P7)}. Finally, several participants reflected on future applications of such systems, such as automating repetitive or less engaging tasks. As one participant suggested: \textit{``I believe it can help reduce the players' burden in games like Genshin Impact. I spent quite a lot of time on daily tasks, but they were repetitive and boring. These monotonous tasks drained my enthusiasm for the game.'' (P8)}.

These qualitative results indicate that players held different expectations for the AI companion depending on their gaming habits and preferred forms of interaction. Although the study modes varied in their level of assistance, participants' experiences were also shaped by perceived agency, interaction cost, and the extent to which the companion felt integrated into the game world.

\section{Discussion}
{The results reveal how different forms of AI assistance and player interaction shaped player experience in our VR game.}
Quantitative workload measures (NASA--TLX) showed that AI-assisted modes reduced players' subjective workload compared with unassisted play. However, game experience measures (MiniPXI) and qualitative feedback indicated that assistance alone was insufficient to sustain a positive gameplay experience when player interaction was absent.

\subsection{Players' Expectations of Interaction and Assistance from AI Companions}

{The results suggest that AI assistance in VR changes not only task difficulty, but also who plans and acts in the virtual environment. Across the four modes, planning, execution, and initiative were distributed differently between the player and the companion. In Agent Play, both planning and execution were transferred to the companion, resulting in the lowest workload but positioning the player largely as an observer rather than an active participant. This interpretation is consistent with Physical Demand showing the largest NASA--TLX subscale effect and with describing Agent Play as ``more like an experience or guidance than a game'' (P8).}

We discuss these findings in relation to the three hypotheses.

\begin{enumerate}[label={}, leftmargin=*]
    \item \textit{\textbf{Hypothesis 1: AI companions reduce players' workload during gameplay.}}

    The quantitative results support this hypothesis. Relative to Single Play, all AI-assisted modes yielded lower overall workload, with Agent Play producing the lowest scores. This pattern suggests that both manual task execution and interaction management contributed to participants' perceived workload. Subscale analyses showed the same general pattern: Agent Play elicited significantly less mental demand, physical demand, and effort than the other modes; Single Play induced higher physical demand and effort than both User Control and Discussion; and Discussion imposed less mental demand than User Control. These findings are consistent with the intended roles of the modes: User Control mainly provided action-execution support while leaving strategy formation to the player, whereas Discussion provided both strategy support and action-execution support.
    
    We observed two workload-related patterns that require further interpretation. First, User Control and Discussion did not differ significantly in overall workload. Although Discussion may have reduced planning demand through shared strategy formation, it also introduced conversational management and latency. User Control, by contrast, minimised dialogue overhead but required players to retain more responsibility for planning. These opposing factors may have offset one another, resulting in similar overall workload. Second, Single Play did not differ significantly from either interactive mode in Mental Demand. Although assistance may reduce planning demands, coordinating with the companion can reintroduce cognitive costs. Participants' qualitative reports, such as comments that Discussion latency complicated play and that acting alone sometimes felt faster, were consistent with this interpretation. 
    {These patterns suggest that workload was shaped not only by the amount of assistance provided, but also by how assistance was requested, negotiated, and enacted during gameplay.}
    
    \item \textit{\textbf{Hypothesis 2: 
    Autonomous assistance without meaningful interaction diminishes players’ gaming experience.
    }}
    
    This hypothesis is supported by both quantitative and qualitative evidence. MiniPXI results showed that Agent Play was rated substantially lower than the interactive modes on key experience constructs such as enjoyment, autonomy, challenge, and immersion. Interview responses similarly attributed this reduced experience to the lack of interaction with the AI companion, with several participants describing Agent Play as less engaging or ``not really a game.''
    
    \item \textit{\textbf{Hypothesis 3: 
    {Moderate assistance improves gaming experience, but too much help degrades it.}
    }}
    
    Our findings provide only partial support for this hypothesis. Although User Control and Discussion produced higher MiniPXI scores than Agent Play, Single Play did not differ from these interactive modes as clearly as the hypothesis would predict. In addition, the modes represented different practical forms of AI companionship, varying not only in assistance level but also in initiative, control, and interaction flow. Thus, the results are better interpreted as differences between interaction designs than as a pure dose-response effect of assistance. Across all four modes, workload levels remained moderate, which may explain why reductions in workload did not consistently translate into improved player experience. Qualitative data further indicated that most participants favoured Discussion Mode or User Control Mode, suggesting that they valued assistance when it preserved interaction and agency. We further categorised participants into four player types based on their self-reported gaming habits and interview responses. We then examined how participants in each category responded to the different AI companion modes. 
    
    Qualitative results revealed recurring preference patterns within these player types:
    \begin{itemize}[leftmargin=*]
        \item \textbf{Task-Oriented Players}: These players focused more on task completion than on conversation, and tended to prefer modes that supported efficient progress, such as User Control. They valued efficiency and were less tolerant of redundant information or interaction overhead during gameplay.
        \item \textbf{Competitive Players}: These players valued dynamic interaction and strategic involvement, but their preferred mode depended on whether they adopted a strategist or executor role during play. 
        \item \textbf{Narrative Players}: These players enjoyed conversational interaction with the AI companion and wanted richer forms of embodiment, such as body language, gaze, or facial expressions. They felt that these interactions could strengthen immersion during gameplay. However, they were also more sensitive to breaks in realism and long response latencies, which could weaken immersion.
        \item \textbf{Social Players}: These players valued the perceived social presence of the AI companion during interaction. They preferred brief but continuous conversational exchanges over long, delayed responses. For these players, responsiveness {appeared to matter} more than response richness, because it helped the companion feel socially present.
    \end{itemize}
    

\end{enumerate}

In summary, participants generally appreciated assistance from the AI companion, but they also valued interactive engagement and a sense of agency. Although assistance reduced workload, fully automated help without meaningful interaction diminished engagement and game experience. Differences across player types further suggest the need for adaptive interaction styles rather than a one-size-fits-all assistance policy.

{
\subsection{Relating the Findings to Prior Work}
Our findings are consistent with, and extend, several strands of prior research. The dissociation we observed between workload and player experience mirrors long-standing evidence that enjoyment in games depends on sustained challenge and perceived competence rather than on minimal effort~\cite{sweetser2005gameflow, ryan2006motivational}: Agent Play removed the demands that made the task meaningful, which is reflected in the sharp drops in the Challenge, Mastery, and Meaning constructs. This pattern also parallels classic results from human--automation research, where high levels of automation reduce engagement and involvement even as they lower workload~\cite{parasuraman2000model, bainbridge1983ironies}. Our results suggest that this tension is amplified in games: whereas in productivity settings task effort is a cost to be minimised, in games effort is part of the value itself, so removing it removes play rather than toil. At the same time, participants explicitly welcomed automation of repetitive ``chore-like'' content, such as daily tasks in \emph{Genshin Impact} (P8), suggesting that the boundary between valued challenge and unwanted toil is content-dependent; this echoes recent work emphasising that AI support in games should be designed around player autonomy~\cite{choong2025support}.

The player-type differences we observed align with established player typologies such as Bartle's player types and Yee's motivational model~\cite{bartle1996hearts, yee2006motivations}. We do not propose our four groups as a new typology; rather, they served as an analytic lens, derived from participants' self-reported habits, for interpreting preferences about AI assistance---a dimension that existing typologies do not directly address. Finally, our results reinforce recent evidence that greater NPC capability does not directly translate into better player experience~\cite{korkiakoski2025empirical}. Participants generally regarded the companion as capable across the assisted modes, yet their experience depended on how assistance was requested, negotiated, and enacted. In the VR setting specifically, the strong effect of gameplay mode on physical demand and participants' descriptions of Agent Play as watching rather than playing suggest that embodied participation is itself a source of engagement---one that assistance can inadvertently remove, and one that is less salient in desktop settings.}

\subsection{Design Guidelines for Future AI NPCs}

Drawing on the quantitative and qualitative findings, we derive the following design guidelines for AI NPCs in VR games:


{First, AI companion interaction should preserve the player’s role as an embodied participant in the virtual world. In VR games, assistance is experienced through spatial action as well as through dialogue. When the companion performs actions that affect the virtual environment they both experience, it can reduce physical and cognitive workload, but it can also make the player feel less involved if planning and execution are fully transferred to the AI. Designers should therefore treat interaction as part of the assistance itself, rather than as an optional conversational layer. Command-based assistance can preserve player initiative while delegating execution, whereas discussion-based assistance can support shared planning when the task justifies the additional interaction cost. Fully autonomous assistance may be useful for repetitive or low-value tasks, but it should be used carefully when the goal is to sustain challenge, autonomy, and immersion.}

Second, AI assistance should be designed around the expectations of different game audiences rather than assuming a uniform form of support. Different games and player communities may expect different forms of interaction from AI companions. For example, progression- or competition-oriented games may benefit from AI companions that prioritise efficient task execution and low-interruption assistance, while narrative or socially immersive games may require richer conversations, emotional interactions, and stronger integration into the game world. Designers should therefore adapt AI companion interaction patterns, such as conversational depth, autonomy, responsiveness, and strategic involvement, to match the intended player experience while reducing repetitive or low-value gameplay demands without diminishing engagement and enjoyment.

Finally, AI companions should maintain social presence through transparent behaviour and well-managed interaction timing.
Players benefit from understanding what the AI companion is doing and when a response can be expected, especially in real-time immersive settings such as VR.
In our study, status cues helped mitigate perceived latency in conversational interaction and supported players' sense of connection with the AI companion.
When low latency cannot be guaranteed, designers can rely on transparency and predictable feedback to preserve the overall gameplay experience.

\subsection{Limitations and Future Work}

Despite the insights gained from this study, several limitations should be acknowledged, representing opportunities for future work.

First, our evaluation was conducted within a single VR puzzle game, which limits the generalisability of the findings across game genres. Although the task allowed us to compare practical forms of AI assistance and interaction under controlled conditions, future studies should examine broader game types, such as action-oriented, narrative-driven, or competitive games, to validate and extend the proposed design guidelines. {The simplicity of the puzzle task also matters for interpretation: several participants noted that the game was easy enough to solve alone, which may have reduced the perceived value of strategic discussion, and preference patterns may differ in games where the companion's strategic input is genuinely needed. In addition, sessions were short and the player-type subgroups were small and unbalanced, so the type-specific preferences reported above require validation with larger samples.}

Second, the four modes were designed as practical interaction configurations rather than a factorial manipulation of individual variables. Future work could use factorial designs to disentangle assistance level, agent initiative, player control, and interaction cost more systematically. {Relatedly, some of the negative feedback on Discussion Mode may reflect the response latency of the current implementation (approximately 3--5\,s) rather than an inherent cost of discussion-based assistance; faster inference could shift the balance between User Control and Discussion. Furthermore, because all assisted modes used the same LLM-driven companion, our study isolates the effect of assistance and interaction modes but not the contribution of the LLM itself: given the constrained action space, a scripted companion could have executed the same commands, and comparing LLM-based with rule-based companions remains an open question for future work. In the future, local model might become viable to run alongside a computer game, and their effectiveness in comparison to remotely-hosted services might represent another avenue for future research.}

Third, the perceptual and expressive capabilities of the NPC remain relatively limited. While the current implementation focused on speech-based interaction and task execution, recent advances in digital human technologies suggest opportunities to incorporate richer multimodal behaviours, such as facial expressions, gaze, body language, and affective responses~\cite{song2023emotional}. Enhancing these channels could further strengthen social presence and improve the perceived intelligence and realism of AI companions.

Fourth, the study examined fixed levels of AI assistance rather than adaptive or personalised behaviours. Although our findings suggest that players differed in their preferences for assistance and interaction, the system did not dynamically adjust its behaviour in response to individual players. Future research could explore adaptive AI companions that infer player preferences over time and adjust the balance between strategy support, action execution, and interaction accordingly.

Finally, our evaluation focused on short-term gameplay within a single session. How player preferences, trust, or reliance on AI companions evolve over repeated sessions remains unexplored. Longitudinal studies should examine how sustained interaction influences player engagement, learning, and perceptions of agency.

\section{Conclusion}

This paper presented an AI NPC framework that supports real-time speech interaction, game-state awareness, and in-game action execution. We deployed the NPC as an AI companion in a VR puzzle game and compared four gameplay modes that varied in assistance, initiative, and interaction flow. Our findings show that AI assistance can reduce players' workload, but fully automated assistance without player interaction can diminish game experience. The results further suggest that players' expectations differ across player types, highlighting the need for AI companions that preserve agency, support meaningful interaction, and adapt to different gameplay preferences.

\bibliographystyle{ACM-Reference-Format}
\bibliography{sample-base}

\appendix

{
\section{Post-hoc Pairwise Comparisons}
\label{app:posthoc}

Tables~\ref{tab:posthoc_tlx} and~\ref{tab:posthoc_pxi} report the Holm-corrected post-hoc pairwise comparisons (paired $t$-tests) for overall NASA--TLX workload and the composite MiniPXI score across the four gameplay modes, including non-significant comparisons. Positive $t$ values indicate higher scores for the first-listed mode. Wilcoxon signed-rank tests with Holm correction, conducted as a nonparametric sensitivity check, produced the same pattern of significant and non-significant comparisons for both measures.

\begin{table}[h]
\centering
\caption{Holm-corrected pairwise comparisons of overall NASA--TLX workload across gameplay modes.}
\label{tab:posthoc_tlx}
\begin{tabular}{lccc}
\toprule
Comparison & $t(23)$ & $p_{\mathrm{Holm}}$ & Cohen's $d$ \\
\midrule
Single Play vs. User Control  & $4.35$  & $<.001$ & $0.95$ \\
Single Play vs. Discussion    & $4.13$  & $.001$  & $1.34$ \\
Single Play vs. Agent Play    & $7.85$  & $<.001$ & $2.30$ \\
User Control vs. Discussion   & $1.01$  & $.321$  & $0.34$ \\
User Control vs. Agent Play   & $4.46$  & $<.001$ & $1.27$ \\
Discussion vs. Agent Play     & $3.42$  & $.005$  & $1.00$ \\
\bottomrule
\end{tabular}
\end{table}

\begin{table}[h]
\centering
\caption{Holm-corrected pairwise comparisons of the composite MiniPXI score across gameplay modes.}
\label{tab:posthoc_pxi}
\begin{tabular}{lccc}
\toprule
Comparison & $t(23)$ & $p_{\mathrm{Holm}}$ & Cohen's $d$ \\
\midrule
Single Play vs. User Control  & $-3.86$ & $.002$  & $-1.21$ \\
Single Play vs. Discussion    & $-1.99$ & $.117$  & $-0.65$ \\
Single Play vs. Agent Play    & $10.86$ & $<.001$ & $3.01$  \\
User Control vs. Discussion   & $1.97$  & $.117$  & $0.47$  \\
User Control vs. Agent Play   & $14.16$ & $<.001$ & $3.96$  \\
Discussion vs. Agent Play     & $11.27$ & $<.001$ & $3.46$  \\
\bottomrule
\end{tabular}
\end{table}

\section{Prompt Templates and Structured Output Schema}
\label{app:prompt}

This appendix reports the prompts used by the Cognitive Layer and the structured output formats parsed by the Executor Layer. Two prompts were used: a command-interpretation prompt in User Control Mode, and a strategy-generation prompt in Discussion Mode. Speech input was transcribed with \texttt{whisper-1} (English), and spoken responses were synthesised with \texttt{tts-1}. Line breaks have been adjusted for column width; the wording is otherwise verbatim.

\subsection{Command Interpretation (User Control Mode)}

{\footnotesize
\begin{verbatim}
You are an NPC in a game where players solve puzzles by
pushing stones to specific locations with a player. The
conversation happens between you and the player.
Your role is to assist the player by interpreting their
commands and executing them using the following functions:
1. Move to a specific position: This function requires one
   parameter:
   - TargetPosition: The position to move to.
     This can be any position or the location of one of the
     following game objects:
     - RedStone, YellowStone, BlueStone, PurpleStone,
       WhiteStone
     - RedCircle, YellowCircle, BlueCircle, PurpleCircle,
       WhiteCircle
     - Player, Wall, Door.

2. Push/move a stone to a specific position: This function
   requires two parameters:
   - TargetStone: The stone to push.
     This can be one of the following game objects:
     - RedStone, YellowStone, BlueStone, PurpleStone,
       WhiteStone
   - TargetPosition: The position to push the stone to.
     This can be any position or the location of one of the
     following game objects:
     - RedStone, YellowStone, BlueStone, PurpleStone,
       WhiteStone
     - RedCircle, YellowCircle, BlueCircle, PurpleCircle,
       WhiteCircle
     - Player, Wall, Door.

IMPORTANT: You must also detect when the player is ending
the conversation (saying goodbye, bye, see you, farewell,
etc.)
If a command cannot be completed using these two functions,
politely inform the user that it is not possible.
If the command can be executed but the parameters are
unclear, ask clarifying questions to gather the required
details.

Your response must include the following parts and be in
JSON format:
1. ExecutionFunction: The function to execute
   (move_to_position, push_stone, or empty/null if no
   command to execute).
2. Parameters: A list of parameters, each with a name and a
   value. Must be exact strings listed above.
3. CommandCompleted: true if the ExecutionFunction is empty
   or if all required parameters are provided; otherwise,
   false.
4. Reply: A short, user-friendly message responding to the
   current user input.
5. ConversationEnd: true if the user is saying
   goodbye/ending the conversation; false otherwise.

Examples:
User: 'move to the red stone' ->
  {"ExecutionFunction": "move_to_position",
   "Parameters": [{"name": "TargetPosition",
                   "value": "RedStone"}],
   "CommandCompleted": true,
   "Reply": "Moving to the Red Stone.",
   "ConversationEnd": false}
User: 'goodbye' ->
  {"ExecutionFunction": null, "Parameters": [],
   "CommandCompleted": true,
   "Reply": "Goodbye! Have a great day!",
   "ConversationEnd": true}
\end{verbatim}
}

\subsection{Strategy Generation (Discussion Mode)}

The placeholders \texttt{\{game\_state\}} and \texttt{\{dialogue\_history\}} are filled at runtime with the serialised game state queried from Unity and the accumulated player--NPC dialogue.

{\footnotesize
\begin{verbatim}
You are a collaborative AI agent working with a human player
to solve a game scenario. The objective of the game is to
open a large gate and cross to the other side of a wall.

### Game Rules:
- The gate is controlled by five magic circles.
- Each magic circle corresponds to a specific magic stone.
- The gate will open only when all five magic circles are
  deactivated simultaneously.
- A magic circle is deactivated when its corresponding magic
  stone is placed on it and it meets specific requirements,
  such as matching the color of the circle.

### Task:
Your role is to analyze the provided game state and propose
a strategy to work with the human player to achieve the
goal. The strategy must include:
1. Clear division of tasks: Specify which tasks the agent
   (you) will perform and which tasks the human will handle.
2. Monitorable objectives: Define clear, actionable
   objectives that can be displayed on the UI for progress
   tracking (e.g., "Move RedStone to RedCircle").
3. Adaptability: Allow for iterative refinement of the
   strategy based on feedback from the human player.

### Communication:
- The player can provide feedback on the proposed strategy.
- Your response should adjust the plan accordingly and
  generate a confirmation statement once both parties agree.

### Output Format:
{
    "agreementStatus": "Pending or Agreed",
    "proposedStrategy": {
        "agentTasks": [
            {"objective": "...",
             "status": "Incomplete",
             "check": {"gameObjectName": "...",
                       "Status": "..."}}
        ],
        "playerTasks": [
            {"objective": "...",
             "status": "Incomplete",
             "check": {"gameObjectName": "...",
                       "Status": "..."}}
        ]
    },
    "responseToPlayer": "..."
}

### Parameter Descriptions:
- agreementStatus: Indicates whether both the agent and the
  player have agreed on the proposed strategy.
  Possible values:
  - "Pending": Agreement not yet reached.
  - "Agreed": Both parties have confirmed the plan.
- proposedStrategy: Contains the division of tasks and
  progress tracking.
  - agentTasks: The tasks assigned to the AI agent.
  - playerTasks: The tasks assigned to the human player.
  Each task includes:
    - objective: A clear, actionable goal
      (e.g., "Move RedStone to RedCircle").
    - status: The completion status of the task.
      Values: "Incomplete" or "Complete".
    - check: Describes how to monitor the task in the game
      engine:
      - GameObject: The name of the game object to track
        (e.g., "RedCircle").
      - Status: The specific property of the game object
        that determines task completion
        (e.g., "activated").
- responseToPlayer: The AI agent's message to the player,
  explaining the strategy or responding to feedback.

### Input:
- Current Game State: {game_state}
- Dialogue History: {dialogue_history}
\end{verbatim}
}

\subsection{Action Validation}

The Executor Layer parses the JSON output and validates it before execution: only the two whitelisted actions (moving to a target and pushing a stone to a target) are executable, and their parameters must exactly match the enumerated game-object identifiers. If the model output contains no executable function or incomplete parameters (\texttt{CommandCompleted = false}), no action is performed and the NPC instead voices a clarifying question. In Discussion Mode, the proposed tasks are tracked against the game state via the \texttt{check} fields, and joint execution starts only after the player confirms the plan (\texttt{agreementStatus = "Agreed"}).
}

\end{document}